\pdfoutput=1
\documentclass[12pt]{article}
\usepackage{amsmath}
\usepackage{graphicx}
\usepackage{natbib}
\usepackage{url} 

\RequirePackage{amsthm,amsmath,amsfonts,amssymb}

\RequirePackage{graphicx,subfigure,latexsym,amssymb,mathtools}

\RequirePackage{float,epsfig,multirow,rotating}
\RequirePackage{upgreek}
\usepackage[noend,ruled,vlined]{algorithm2e}

\theoremstyle{plain}

\newtheorem{theorem}{Theorem}[section]

\theoremstyle{remark}
\newtheorem{definition}[theorem]{Definition}

\numberwithin{equation}{section}
\theoremstyle{plain}
\DeclareMathOperator\erf{erf}
\DeclareMathOperator\erfc{erfc}
\newtheorem{remark}{Remark}[section]

\newcommand{\bSigma}{\boldsymbol{\Sigma}}

\newcommand{\bV}{\boldsymbol{V}}

\usepackage{authblk}

\newcommand{\blind}{0}

\begin{document}

\def\spacingset#1{\renewcommand{\baselinestretch}%
{#1}\small\normalsize} \spacingset{1}


\if0\blind
{
  \title{\bf Identifying critical residues of a protein using meaningfully-thresholded Random Geometric Graphs}

\author[1]{Chuqiao Zhang\thanks{We gratefully acknowledge the invaluable insights of Prof. Gesine Reinert towards the development of the thresholding technique that we report here.}\thanks{For MD simulations of Skp1, we made use of time on HPC granted via the UK High-End Computing Consortium for Biomolecular Simulation, HECBioSim (http://hecbiosim.ac.uk), supported by EPSRC (grant no. EP/X035603/1).}}
\author[1,2]{Sarath Chandra Dantu}
\author[3]{Debarghya Mitra}
\author[4]{Dalia Chakrabarty}

\affil[1]{Department of Mathematics, Brunel University London, Uxbridge UB8 3PH, UK}
\affil[2]{The Thomas Young Centre for Theory and Simulation of Materials, London SW7 2AZ, UK}
\affil[3]{Department of Bioscience and Bioengineering, Indian Institute of Technology, Mumbai, India}
\affil[4]{Department of Mathematics, University of York, York YO10 5DD, UK}
  
  \maketitle
} \fi

\if1\blind
{
  \bigskip
  \bigskip
  \bigskip
  \begin{center}
    {\LARGE\bf Title}
\end{center}
  \medskip
} \fi

\bigskip
\begin{abstract}
Identification of critical residues of a protein is actively pursued, since such residues are essential for protein function. We present three ways of recognising critical residues of an example protein, the evolution of which is tracked via molecular dynamical simulations. Our methods are based on learning a Random Geometric Graph (RGG) variable, where the state variable of each of 156 residues, is attached to a node of this graph, with the RGG learnt using the matrix of correlations between state variables of each residue-pair. Given the categorical nature of the state variable, correlation between a residue pair is computed using Cramer's V. We advance an organic thresholding to learn an RGG, and compare results against extant thresholding techniques, when parametrising criticality as the nodal degree in the learnt RGG. Secondly, we develop a criticality measure by ranking the computed differences between the posterior probability of the full graph variable defined on all 156 residues, and that of the graph with all but one residue omitted. A third parametrisation of criticality informs on the dynamical variation of nodal degrees as the protein evolves during the simulation. Finally, we compare results obtained with the three distinct criticality parameters, against experimentally-ascertained critical residues.
\end{abstract}

\noindent%
{\it Keywords:}  
protein design; Random Geometric Graphs; thresholding; rejection sampling; degree distribution

\spacingset{1.75} 

\section{Introduction}
\label{sec:intro}

Proteins are biopolymers that
contribute intimately to almost all cellular processes, such as biochemical catalysis; signal transduction; immunity, etc. 
A residue is a
fundamental unit of a biopolymer, and certain residues - referred to as critical residues - are essential for the protein to function, contributing to
the determination of pathogenic effects of mutations; to rational protein engineering; and to targetted drug design. Hence, experimental and computational techniques are rigorously pursuing the identification of critical residues in a protein \citep{Elcock2001,delsol2003}.

Several computational strategies have been developed to identify these
residues, including analysis of rich protein sequence alignments \citep{Thomas1991,sander1991,Shenkin1991, Capra2007, Hopf2017}, via
network-driven approaches that use molecular dynamical simulations
\citep{Yehorova2024, xu2025}.  Sequence conservation analysis refers to the monitoring of invariance of the sequence of amino acids during the
time interval over which evolution of the protein is noted. Measures
of invariance based on Shannon entropy have been forwarded by \cite{Thomas1991,sander1991,Shenkin1991}. Here,
critical residues exhibit low entropy, indicating high functional
importance.  \cite{Capra2007} used an extension of Shannon entropy,
namely, the Jensen-Shannon Divergence, while \cite{Hopf2017}
introduced an unsupervised probabilistic method using evolutionary
data, effectively capturing residue-residue dependencies, to predict
pathogenic effects of mutation.

Another popular approach is to identify critical residues based on their contribution to protein stability and binding energetics. These thermodynamic methods assume that functionally important residues minimise the free energy of a protein's native state and facilitate conformational transitions \citep{Elcock2001, Su2011}. However, there are some limitations of these approaches in the current studies. Sequence
conservation-based methods fail to account for protein structure and dynamics - some highly conserved residues may not be functionally significant, while functionally important residues that evolved for specific structural or allosteric roles may not be strictly conserved. Again, thermodynamic methods provide high-resolution functional insights, but they are computationally expensive and often require highly accurate protein structures. Additionally, they do not directly capture network-wide residue interactions beyond local energetic contributions.

To address these limitations, network-based models have emerged as a powerful alternative for analysing residue importance in proteins.
Recent MD-based strategies alleviate this problem by constructing rigid graphs from analysis of interaction networks or ranked or linear correlation matrices built from analysis of structural dynamics and then invoke centrality measures to identify important residues. While this has allowed recovering long-range interaction networks, this still relies on community detection analysis and centrality measures on static graphs \citep{Doncheva2012,Haspel2017,Kantelis2022,Osuna2020,Gerek2013,Yehorova2024,xu2025}.

\cite{giles} proposed a Soft Random Geometric Graph (SRGG) of a protein, which assigns probabilistic edge weights instead of binary interactions. In our work, we introduce a new Random Geometric Graph (RGG) drawn in a probabilistic metric space, to formulate multiple parametrisations of criticality. We undertake probablistic mechanistic learning of this graph variable given simulated data on states attained by residues of a protein, and apply it to motivate multiple parametrisations of temporal fidelity of secondary structure state of individual residues.

We illustrate these techniques for identification of critical residues in
the S-phase kinase protein {\textit{Skp1}}, using data obtained from molecular dynamical (MD) simulations of this protein. Section~1 and Section~2 of 
of the Supplement respectively present details of these simulations and the biological importance of {\textit{Skp1}}. 
Using such simulated data on the categorical state variable, we compute the inter-residue correlation matrix of this protein, and thereafter, employ the computed correlation matrix to construct a realisation of the RGG variable. To any node of the learnt graph, is associated the random variable that is the state attained by a residue at a time point, where said time point lies within the time interval, during which, outputs are recorded in the simulation. In the learnt realisation of the random graph variable, edges are learnt to exist between a pair of nodes (i.e. residues), with a probability, given the inter-residue correlation matrix computed given the simulated data on states attained by residues. Probability for the edge variable that joins a nodal pair, informs on the interaction between the pair of residues that respectively sit at these two nodes.

\section{Methodology}
\subsection{Data}
\label{sec:data}

\begin{figure}[H]
     \centering
     \includegraphics[width=0.4\textwidth]{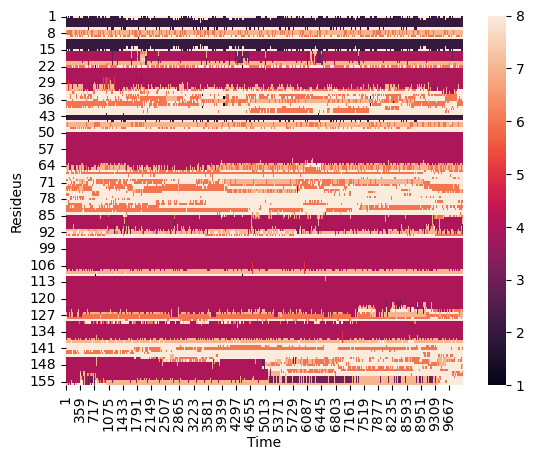}

     \caption{Figure showing a heat map of the simulated data ${\bf D}$ that depicts the state attained by the $k$-th residue at the $t$-th time point, for $t=1,\ldots, N_T=10006$ and $k=1,\ldots,156$. The categorical state variable that takes values at eight levels, is denoted one of eight shades, each of which represents a configuration of the residue.}
\label{fig:1}
\end{figure}

Section~1 of the Supplement provides details on the MD simulation setup and the secondary structure data from each residue. This simulation trackes the dynamic evolution of 158 residues of {\it{Skp1}}, over the time interval $[0,N_T]$,
where the time point $N_T=1000.6$ is in nanoseconds (ns), and the $i$-th row of the simulated dataset reports on the evolution during the temporal interval $[0.1(i-1), 0.1i]$ ns, $i=1,2,\ldots,10006$. 
There are eight distinct states that any residue can attain in the simulation, (Section~1 of the Supplement), such that (s.t.) the state variable is a categorical one, taking values at eight levels. Within the simulation output, each state is depicted in one of eight colours. In Figure~\ref{fig:1}, we present the temporal evolution of the 158 residues of the protein {\it{Skp1}}, with the states depicted in distinct colours.

States attained by all residues at each time point - at which an output is recorded in the simulation - forms a distinct row of the simulated dataset ${\bf D}$, while states attained by each residue forms a column of this dataset. There are 10006 rows of this dataset, and though
158 residues of {\it{Skp1}} were included in the simulations, the first and last residues did not undergo any change during the time interval tracked within the simulations, and were therefore ignored from the considered dataset ${\bf D}$. Thus, we use 156 residues, and denote these with the index variable $K$ that takes the value $k\in\{1,\ldots,156\}$, in our definition. We refer to the state attained by the $k$-th residue at the $t$-th time point as $x_{k}^{(t)}$. Then $x_{k}^{(t)}$ is the value of the categorical state variable $X_k\in{\cal X}$ that is attained by the $k$-th residue at time $T=t$, where $k=1,\ldots,156$. Then $X_k$ takes values at eight string-valued levels. Here, $t$ is the value of the variable $T\in{\cal T}\subset{\mathbb N}$ that represents the time point at which the MD simulations output values of the state attained by each residue of {\it{Skp1}}, s.t. $t=1,2,\ldots,N_T=10006$. Thus, ${\bf D}=\{x_{k}^{(t)}\}_{k=1; t=1}^{156; N_T}$.



\subsection{Random Geometric Graph variable}

We learn a realisation
of a Random Geometric Graph (RGG) variable \citep{giles, Penrose:2016, plos23,joe24}, of dataset ${\bf D}$, where said RGG is drawn in
the probabilistic metric space \citep{Menger1942,sklar}.
\begin{definition}
Similar to a non-negative distance function that can be computed between any two
points in a metric space, in a probabilistic metric space $\{{\Omega}, d(\cdot, \cdot),
\Delta(\cdot,\cdot)\}$, we can
compute the distance function $d(x_i, x_j)$ between the two points
$X_i\in{\cal X}$ and $X_j\in{\cal X}$, with this distance function given
  by the cumulative distribution function (i.e. {\it{cdf}}) $F_{S(X_i, X_j)}(s)$ of the
  variable $S(\cdot, \cdot)$ that is a known function of
  $X_i$ and $X_j$, s.t. $F_{S(\cdot,\cdot)}(\cdot)$ enjoys positive
  support, $\forall X_i,X_j\in{\cal X}$, and $s(x_i,x_j)\geq 0$ is a
  value of $S(X_i,X_j)$. Here the mapping $S:{\cal X}\times{\cal X}\longrightarrow{\Omega}$ is defined below. Thus,
  $d(x_i,x_j)=F_{S(X_i,X_j)}(s(x_i,x_j))$, and the triangle function
  $\Delta(\cdot,\cdot)$ is s.t.  $\Delta(F_{S(X_i,X_j)}(\cdot),
  F_{S(X_j,X_k)}(\cdot)) \geq F_{S(X_i,X_k)}(\cdot)$, $\forall
  X_i,X_j,X_k\in{\cal X}$.
  \end{definition}

The RGG that we construct in the probabilistic metric space, has $p=156$ nodes in the vertex set $\bV = \{1,2,\ldots,p\}$, with the random variable $X_i$ attached to the $i$-th node, $\forall i\in\bV$, and the edge variable between the $i$-th and $j$-th nodes denoted $G_{i,j}$ s.t. $G_{i,j}=g_{i,j}\in\{0,1\}$. Then as in any RGG, here too, the edge between the $i$-th and $j$-th nodes exists, (i.e. $g_{i,j}=1$ instead of 0), as long as the mutual inter-nodal distance $d(x_i,x_j)$ exceeds a cut-off $\tau$. Since the inter-nodal distance of this RGG is a probability - i.e. $d(x_i,x_j) = F_{S(X_i,X_j)}(s(x_1,x_j))$ - the cut-off $\tau$ is a probability, i.e. $\tau\in[0,1]$.

Here $F_{S(\cdot,\cdot)}(\cdot)$ is a {\it{cdf}} with positive support, where the function $S(x_i,x_j)$ measures the ``disparity'' between the absolute correlation of $X_i$ and $X_j$, and the connectivity of the $i$-th and $j$-th nodes, i.e. $S_{i,j} \equiv S(x_i,x_j) := \vert(G_{i,j} - \vert R_{i,j}\vert\vert$, where the correlation matrix $\bSigma =[\vert corr(X_i,X_j)\vert]$, with the correlation between $X_i$ and $X_j$ given as $corr(X_i,X_j) \equiv R_{i,j}$, for $i,j\in\bV$. So the disparity $S_{i,j} \equiv S(x_i,x_j) := \vert(G_{i,j} - \vert R_{i,j}\vert\vert\in[0,1]$ s.t.
$S:{\cal X}\times{\cal X}\longrightarrow[0,1]=\Omega$. Then $S_{i,j}$ is lowest in value, if $G_{i,j}=1$ (or $G_{i,j} =0$), when $\vert corr(X_i, X_j) \vert$=1 (or $corr(X_i,X_j)=0$).
It can be shown that 
\begin{equation}
  d(x_i,x_j) = F_{S_{i,j}}(s_{i,j}) = K\displaystyle{\left[\sqrt{\frac{2}{\pi}}(e^{-{s_{i,j}^2/2}}) - s_{i,j} \erf(s_{i,j}/ \sqrt{2})\right]},
  \label{sec:doori}
  \end{equation}
is a distance function in ${\cal X}$, where $s_{i,j}$ is the value of $S_{i,j} = \vert(G_{i,j} - \vert R_{i,j}\vert\vert$. Given that the distance function is a {\it{cdf}}, the constant $K$ in its definition is $[\erf(1/\sqrt{2}) + \sqrt{2/e\pi} - \sqrt{2/\pi}]^{-1}$.

We wish to compute the posterior probability of the edge variable $G_{i,j}$, conditional on the absolute correlation betweeb $X_i$ and $X_j$.
For data ${\bf D}$ that comprises $N_T$ observations of the random variable $X_i$ in its $i$-th column, $\forall in \in\bV$, let the inter-column correlation matrix be $\bSigma$, i.e. $\bSigma = [\vert corr(X_i,X_j)\vert]$, with $corr(X_i,X_j)=R_{i,j}, \forall i < j; j\in\bV$. The edge variable $G_{i,j}$ of this RGG is then learnt, conditional on the correlation $R_{i,j}$, s.t. we can define a posterior probability $m(G_{i,j}=g_{i,j}\vert \vert R_{i,j}\vert=\rho_{i,j})$. One possible definition of the {\it{pdf}} of the observable $\vert R_{i,j}\vert$ is ${\cal N}(g_{i,j}, \nu)$, when the edge variable $G_{i,j}$ attains the value $g_{i,j}$, and the variance variable $\sigma^2$ attains the value $\nu$. In other words, $f_{\vert R_{i,j}\vert}(\rho_{i,j}\vert g_{i,j},\nu) = {\cal N}(g_{i,j},\nu)$. When this density is computed at $\rho_{i,j} =g_{i,j}$, the density is a maximum, and the density computed at $\rho_{i,j} =1 - g_{i,j}$ is a minimum - given $G_{i,j}=g_{i,j}\in\{0,1\}$, at all values of $\sigma^2$.
\begin{definition}
\label{defn:karon}
  Using a Bernoulli(0.5) prior on $G_{i,j}$ and Uniform[0,1] prior on $\sigma^2$, the joint posterior density of $G_{i,j}$ and $\sigma^2$, computed at $G_{i,j}=g_{i,j}$ and $\sigma^2=\nu$, conditional on $\vert R_{i,j}\vert=\rho_{i,j}$, is proportional to ${\cal N}(\rho_{i,j}, \nu)$. Marginalising $\nu$ out of this joint posterior then leads to the edge marginal posterior $m(G_{i,j}=g_{i,j}\vert \rho_{i,j}) \propto \sqrt{2/\pi}e^{-s_{i,j}^2/2} - s_{i,j} \erfc (s_{i,j}/\sqrt{2})$, where we recall that $s_{i,j}$ is the value of $S_{i,j}=\vert G_{i,j} - \vert R_{i,j}\vert\vert$, s.t. $s_{i,j} = \vert g_{i,j} - \rho_{i,j}\vert$. Here $\erfc(\cdot)$ is the complimentary error function.
\end{definition}
Indeed, this marginal posterior of the $i,j$-th edge, given $\vert R_{i,j}\vert = \rho_{i,j}$, decreases (or increases) as the distance $d(x_i, x_j)=\Pr(S_{i_j} \leq s_{i,j})$ increases (or decreases), where the distance function $d(\cdot,\cdot)$ is given in Equation~\ref{sec:doori}. So as in an RGG, when an edge is modelled to exist only if the inter-nodal distance falls short of a pre-chosen threshold probability $\tau$, in the graph that we learn, we set: $g_{i,j}=1\iff m(G_{i,j}=1\vert \rho_{i,j}) > \tau$. Although the RGG variable indicates its $\tau$ dependence, the
posterior of this graph variable defined as $\prod_{i,j\in{\bV}; i< j}
m(G_{i,j}=g_{i,j}\vert \rho_{i,j})$, is independent of the pre-chosen
probability cut-off $\tau$.

\begin{definition}
  \label{defn:rgg_post}
The posterior of the RGG variable ${\cal G}_{\bV, m}(\tau, {\bSigma})$ that
is learnt using the edge posterior $m(\cdot\vert \cdot)$, given the
known inter-column correlation matrix $\bSigma=[\rho_{i,j}]$ of data ${\bf D}$, is $\pi({\cal G}_{\bV, m}(\tau,\bSigma)\vert \{\rho_{i,j}\}_{i,j\in\bV}) = \prod_{i,j\in{\bV}; i< j} m(G_{i,j}=g_{i,j}\vert \rho_{i,j})$,
  where the RGG is constructed without self-loops, and the edges are
  included independently of each other. Here $m(\cdot\vert\cdot)$ is defined in Definition~\ref{defn:karon}. 
\end{definition}


\subsection{Implementation of graph learning}
We learn the RGG of data ${\bf D}$ as follows.

\begin{enumerate}
    \item[---] Learn the RGG ${\cal G}_{\bV,m}(\tau, \bSigma)$, defined on the vertex set $\bV := \{1,\ldots, p\}, p=156$, given inter-residue correlation matrix $\bSigma$ of dataset ${\bf D}$.
\item[---] In fact, the unbiased estimate of $\vert corr(X_i,X_j)\vert$ $\forall i < j; j\in\bV$, using the $N_T$-sized sample of state $X_i$ attained by the $i$-th residue, $\forall i=1,\ldots,156$.
\item[---] Given this estimated $156\times 156$-dimensional inter-residues matrix $\bSigma$ at hand, we then compute the closed-form posterior probability of any edge of the graph variable (Definition~\ref{defn:rgg_post}). s.t. we can sample from this posterior.
\item[---] We perform Rejection Sampling with proposal density $q_{i,j}$, to draw $N$ samples of the edge variable $G_{i,j}$, s.t. in the $r$-th sample, the binary edge variable $G_{i,j} = g_{i,j}^{(r)}$. We choose $M > 1$ s.t. ${m(g^{(r)}_{i,j}\vert \rho_{i,j})} < {M q_{i,j}}$, 
and accept the proposed edge if and only if ${m(g^{(r)}_{i,j}\vert \rho_{i,j})} /{M q_{i,j}} \geq u$, where $U=u$ and $U \sim \text{Uniform}[0,1]$.
\item[---] The relative frequency of the edge variable $G_{i,j}$ in the generated sample 
$\{g_{i,j}^{(1)}, g_{i,j}^{(2)}, \ldots, g_{i,j}^{(N)}\}$, is $f_{i,j} := {{\sum_{n=1}^{N}
g_{i,j}^{(n)}}/{N}}$. Then in the graph of data ${\bf D}$, $g_{i,j}=1\iff f_{i,j} > \tau$, $\forall i <j; j\in\bV$. 
\end{enumerate}

To undertake automated identification of the $n$-most “critical” residues in a protein, using its simulated data ${\bf D}$ on states attained by its residues, we next propose three distinct parameters.

\subsection{Parametrisation of functional instability of a residue}
We define the parameter $\delta_c$, as a difference between
the posterior probability of the RGG variable, given the data ${\bf D}$, (stated in Definition~\ref{defn:rgg_post}), and the posterior probability of the RGG variable given the data ${\bf D}_{-c}$
that consists of values of the state attained by all residues of
{\it{Skp1}} at time points $1,2,\ldots,N_T$, exccept for the $c$-th
residue. Here $c=1,\ldots,156$. Thus, ${\bf D}_{-c}$ is a dataset
$\{x_{k}^{(t)}\}_{k=1, k\neq c; t=1}^{156; N_T}$, and let the inter-residue correlation matrix of this dataset be denoted
$\bSigma_{-c}$.

\begin{algorithm}[H]
\SetAlgoLined
\SetKw{KwBy}{increment by}
\SetKwInOut{Input}{Input}
\SetKwInOut{Output}{Output}
\tcc{Learning RGG of data ${\bf D}$.}
\text{States attained by $c$-th residue at $t=1,...,N_T=10006$, $\forall c=1,\ldots,156$,}
\text{populate the $N_T\times 156$-dimensional dataset ${\bf D}$}

\text{Estimate $156\times 156$-dimensional inter-column correlation matrix $\bSigma=[\rho_{i,j}]$.}
  
    \For{$i\gets1$ \KwTo $156$ \KwBy $1$,}{
        \For{$j\gets1$ \KwTo $156$ \KwBy $1$,}{
            \text{Given $i,j$-th element of $\bSigma$, sample
              $g_{i,j}$ from edge marginal}\\
            \text{$m(G_{i,j}=g_{i,j}\vert \rho_{i,j})$ given in Definition~\ref{defn:karon}}\\

            \text{Compute posterior probability $\pi({\cal G}_{\bV, m}(\tau,\bSigma)\vert \{\rho_{i,j}\}_{i< j; j\in\bV})$ of random graph} 
            \text{variable ${\cal G}_{\bV, m}(\tau, {\bSigma})$}

            \For{$r\gets1$ \KwTo $N$ \KwBy $1$,}{
            \text{Sample from posterior probability ${\cal G}_{\bV, m}(\tau, {\bSigma})$ using Rejection Sampling,} to generate sample $\{g_{i,j}^{(1)}, g_{i,j}^{(2)}, \ldots, g_{i,j}^{(N)}\}$ for edge variable $G_{i,j}$
            }
        \text{Compute relative frequency $f_{i,j}:={{\sum_{n=1}^{N}
g_{i,j}^{(n)}}/{N}}$}. $G_{i,j}=1 \iff f_{i,j} > \tau$.
        }
        }
\caption{Algorithm for implementation of RGG}
\label{algo:implementation}
\end{algorithm}


\begin{definition}
  \label{defn:delta}
We learn the RGG
variable ${\cal G}_{\bV_{-c},m}(\tau, \bSigma_{-c})$ defined on the vertex set $\bV_{-c} := \{1,\ldots, c-1, c+1, \ldots, p\}$, given inter-column correlations $\{\rho_{i,j}\}_{i,j\in\bV_c}$ of data ${\bf D}_{-c}$, $\forall c\in\{1,\ldots,156\}$,
and compute $\delta_c$ as:
$$\delta_c := \pi({\cal G}_{\bV_{-c}, m}(\tau, \bSigma_{-c}\vert \{\rho_{-c}\}_{i,j\in\bV_c})
  - \pi({\cal G}_{\bV, m}(\tau, \bSigma)\vert \{\rho_{i,j}\}_{i,j\in\bV}),$$
    $\forall c=1,\ldots,p=156$.
Here, $\bSigma_{-c}$ is the inter-residue correlation matrix of dataset ${\bf D}_{-c}$, s.t. the $e,f$-th element of $\bSigma_{-c}$ is $\vert corr(X_e, X_f)\vert$, where $e,f\in \bV_{-c}$.    
Then these computed values $\delta_1,\delta_2,\ldots,\delta_p$ are sorted in ascending order, s.t. higher is the value of $\delta_c$, the more critical is the $c$-th residue to the protein.
\end{definition}
\begin{remark}    
  If the $c$-
  th residue provides a high $\delta_c$, then it implies that the more frequent bonding and unbonding of this critical residue with other residues - i.e. the instabiliy of this $c$-th residue - leads to the RGG variable ${\cal G}_{\bV_{-c}, m}(\tau, \bSigma_{-c})$ to be more compatible with the correlation $\bSigma_{-c}$, than the RGG ${\cal G}_{\bV, m}(\tau, \bSigma)$ is with $\bSigma$. Thus, $\delta_c$ parametrises the functional instability of the $c$-th residue. $\delta_c$ is not normalised, s.t. $\delta_c\in{\mathbb R}$.
\end{remark}

One advantage of using the $\delta_{\cdot}$ parameter for criticality identification is that it is independent of the choice of the cut-off probability $\tau$, unlike the other parameters we introduce below. However, computation of all - i.e. the $p$ number of - $\delta_{\cdot}$ parameters is required, in order to ascertain the ranking of residues by criticality. Since $p$ is typically high (=156 for {\it{Skp1}}) this implies that we need to learn $p+1$ realisations of RGGs, given the inter-residue correlation matrices $\bSigma$ and $\bSigma_{-1},\ldots,\bSigma_{-p}$, that we will have to compute, using the datasets ${\bf D}$ and ${\bf D}_{-1}, \ldots, {\bf D}_{-p}$, respectively. This is time and resource-intensive. Hence the disadvantage of this approach for identifying the $n$-most critical residues of a given protein, where $n\in{\mathbb N}$ is set by the practitioner, with $n < p$.

\subsection{Choice of the cut-off $\tau$}
\begin{remark}
  The choice of the cut-off $\tau$ affects the degree of the $c$-th node in the RGG learnt with the data ${\bf D}$ (or with any
  partition of data ${\bf D}$).
  Here $c\in\bV$.
We need to check the effect of $\tau$ on any parametrisation of criticality that uses the nodal degree as an underlying measure. We will discuss two such parameters below.
It is relevant that such parameters be computed at a ``meaningful'' $\tau$. By ``meaningful'' here, we imply a $\tau$ that is not chosen arbitrarily, but instead produces the ``most robust RGG'' typically, (and on rare occasions, the ``most sensitive RGG''), where the considered ``robustness'' (or ``sensitivity'') is to changes in the thresholding, i.e. to changes in $\tau$.
\end{remark}

Indeed, there are various existing studies on the determination of thresholds that are relevant to network construction \citep{Freeman2007,Bassett2008,Drakesmith2015}. One of the simplest and computationally efficient ways is to choose a constant threshold or cutoff, when constrcuting networks \citep{Freeman2007}. \cite{Ala2008} retained only the top $k$ percent of the edges in terms of correlation, which is useful in maintaining network sparsity. Such a choice of imposing a fixed cutoff is however unsatisfactory, since the choice of the threshold is subjective, and can lead to inconsistencies across datasets. Also, fixed thresholds may oversimplify complex network structures, such as biological networks, in which biologically-relevant interactions do not necessarily have high correlations, but can still significantly contribute to network functionality.
\cite{Perkins2009} tackled the issue of determining appropriate thresholds when constructing gene co-expression networks using spectral graph theory. However, this technique is computationally-intensive, rendering it infeasible for application, when large-scale networks are considered. \cite{Theis2023} evaluated thresholding techniques by introducing the Objective Function Threshold (OFT) method, which determines optimal thresholds for weighted brain networks, by comparing several graph metrics, including density; transitivity and clustering coefficient of the network; and the Characteristic Path Length (CPL). Here CPL \citep{watts1998}, is the average number of steps (or edges) needed to travel between any two nodes in the network, and is identified to be optimal. However, CPL is affected by isolated nodes or disconnected components in the network, and thresholding using this technique can be misleading in large networks, since the frequency of longer paths increases with network size.

Distinguished from existent approaches in the literature, we seek a data-driven thresholding method, such that the threshold selection is generic, yet relevant to the context at hand, and crucially, the choice is motivated by the nature of the graph that is being learnt for a given dataset. We concur with \cite{cogent_1,cogent_2} that consistency of the realised network/graph to changes in the thresholding is important. \cite{cogent_1,cogent_2} advance the technique ``COGENT'' to identify the most consistent network at a chosen threshold $\omega$, by comparing networks generated by overlapping subsets of the given data, with similarity quantified using metrics, such as the overlap edge set and the correlation between node degrees. Here this pre-chosen threshold $\omega$ contributes to network construction by including only those edges s.t. the correlation between the straddling nodal-pair is within the top $(1-\omega)\%$ of correlation values. Our pursuit of the most robust RGG, (defined on a given vertex set and given an inter-residue correlation), is similar in spirit to the most consistent network, but organically defined by our treatment of the graph as a random variable - the threshold value $\tau$, at which rate of change of the posterior of the learnt RGG is least, is an optimal $\tau$.

\begin{definition}
\label{defn:most}
  The RGG ${\cal G}_{\bV, m}(\tau, {\bSigma})$, the posterior of which (as in Definition~\ref{defn:rgg_post}) changes the least with changes in $\tau$, is referred to as the most robust RGG, (or strictly speaking, the most robust realisation of the RGG variable). The $\tau$ at which the most robust RGG is attained is denoted $\tau_{min}$. 

  Again, the $\tau$ at which the posterior of the RGG variable changes the most, is a ``meaningful $\tau$'', and the corresponding RGG realisation is then the one that is most sensitive to changes in $\tau$. This $\tau$ is denoted $\tau_{max}$.
\end{definition}

We compute the value of the (logarithm of the) posterior $\pi({\cal
  G}_{\bV, m}(\tau,\bSigma \vert 
  \{\rho_{i,j}\}_{i<j; j\in\bV})$ of
  the RGG variable, (conditional on the inter-residue correlation
  matrix $\bSigma$ of data ${\bf D}$), at pre-selected values of
  $\tau\in[0,1]$. Then we approximate the slope of this log posterior
  by differencing between the log posterior generated at neighbouring
  values of $\tau$. We denote this slope $\gamma(\tau) :=
  d\log[\pi({\cal G}_{\bV, m}(\tau,\bSigma \vert \{\rho_{i,j}\}_{i<j;
    j\in\bV})]/d\tau$. Approximating $\gamma(\tau)$ with the
    differencing leads to a noisy identification of itself. We
    identify the values of $\tau$ at which this slope is a maximum,
    indicating the values of $\tau$ at which the rate of change of the
    logarithm of the posterior of the RGG variable, with changes in
    $\tau$, is maximum. Such values of $\tau$ are then our
    empirically-identified $\tau_{max}$. On the other hand, the values of
    $\tau$ at which $\gamma(\tau)$ display minima, are identified as
    those, at which the learnt realisation of the RGG variable changes
    least with a given change in the value of $\tau$. Then such
    empirically-noted minima of the slope of the log of the RGG
    posterior with $\tau$, correspond to $\tau=\tau_{min}$.


We will compare the results of our thresholding using $\tau$ against the thresholding by \cite{cogent_1} using the method COGENT, which aims to find the most consistent network with respctive to the used thresholds.


\subsection{Parametrisation of temporal variation of degree distribution}
The second parameter that we forward to inform on the critical residues, is the standard deviation of the sample of temporally-local values of the degree of each of the $p$ nodes of the RGG that is learnt with dataset ${\bf D}$. The motivation is that, as the protein evolves, a critical residue manifests comparatively higher non-uniformity in its degree, since it initiates and terminates interactions with other residues of the protein more frequently, than non-critical residues. Thus, the standard deviation $\eta_c$ of the temporal distribution of the degree of the $c$-th node, will be higher than the value of $\eta_{c^{/}}$, if the $c$-th residue is more critical than the $c^{/}$-th residue, $\forall c,c^{/}\in\bV$. $\eta$ can then be acknowledged as a dynamic parameter.

\begin{definition}
  We partition the dataset ${\bf D}$ into $N_b$ blocks of $N_{\eta}$ rows each, s.t. the $b$-th block is the $N_{\eta}\times p$-dimensional matrix ${\bf D}_{b}$, elements of which are values of the state variable of each of the $p$ residues of the protein, over $N_{\eta}$ time points, $\forall b=1,\ldots,N_b := N_T/N_{\eta}$. Let the inter-residue correlation matrix of dataset ${\bf D}_b$ be $\bSigma_b=[\rho_{i,j}]$, $\forall i,j\in\bV$. RGG ${\cal G}_{\bV, m}(\tau, {\bSigma_b})$ is learnt at a pre-chosen cut-off probability $\tau$, $\forall b=1,\ldots,N_b$, using dataset ${\bf D}_b$, s.t. in this graph, $g_{i,j} =1$ if the edge posterior $m(G_{i,j}=1\vert \rho_{i,j}) > \tau$. 

Let $\eta_c^{(b)}$ be number of edges connected to the $c$-th node in the graph ${\cal G}_{m,\bV}(\tau, \bSigma_b)$ learnt at a chosen value of $\tau$. Thus, 
  $$\eta_c^{(b)} := \vert \{(c,i): g_{c,i}=1, i\neq c, i\in\bV\}\vert, \forall c\in{\bV}.$$
  Then the unbiased estimate of standard deviation $\eta_c$ of sample $\{\eta_c^{(1)}, \eta_c^{(2)}, \ldots,\eta_c^{(N_b)}\}$ is $$\eta_c = \sum_{i=1; i\neq c}^{N_b} \sqrt{(\eta_c^{(i)} - {\bar{\eta}}_c)^2/(N_b-1)}, \text{ where }{\bar{\eta}}_c = \sum_{i=1; i\neq c}^{N_b} \eta_c^{(i)}/N_b.$$
Then $\eta_1,\ldots,\eta_p$ for $p=1,...,156$, is sorted and the $n$-highest values of these parameters correspond to the $n$-most critical residues in the protein.    
\end{definition}

Therefore $\eta_c$ computed using RGGs realised at a given $\tau$, is a reliable indicator of criticality of the $c$-th residue, $\forall c\in\bV$, based on the guiding principle that as the protein evolved, interaction between a critical residue and other residues start and end comparatively more abruptly and frequently - than that for a non-critical residue. As a result, bigger changes in the temporal distribution of degree of a node, (over another node), are indicative of comparatively higher criticality of the residue that corresponds with the former node. We typically find the variation in the degree distribution of the most-robust RGGs learnt across times, i.e. we choose to work with $\tau=\tau_{min}$.

\subsection{Parametrising the degree of individual nodes in the full graph}
While $\eta_c$ informs on criticality of the $c$-th residue, via the temporal evolution of the degree of the $c$-th node, its degree computed over the entire time interval $[0,N_T]$ - denoted $\beta_c$ - could also inform on the criticality of this residue. 
\begin{definition}
  $\beta_c := \vert \{(c,i): g_{c,i}=1, i\neq c, i\in\bV\}\vert$, $\forall c\in{\bV}$, where $G_{c,i}$ is the edge variable straddled by the $c$-th and $i$-th nodes of the RGG ${\cal G}_{\bV, m}(\tau, {\bSigma})$.
\end{definition}
Then $\beta_c$ is affected by the pre-chosen cut-off probability $\tau$. In our work, we choose to work with the $\beta_{\cdot}$ parameter that is computed in the most robust realisation of the RGG variable - to changes in $\tau$ - i.e. the RGG defined at $\tau=\tau_{min}$.
\begin{remark}
If the $c$-th residue is more critical than the $c^{/}$-th residue, then the $c$-th residue connects and disconnects with other residues more frequently than the $c^{/}$ residue, during the evolution (of the protein) noted over the whole time interval $[0,N_T]$. Thus, the degree of the $c$-th residue, noted in the data collected over this time period, is lower than the degree of the $c^{/}$-th residue, i.e. then $\beta_c < \beta_{c^{/}}$ in general. Here $c\neq c^{/}; c,c^{/}\in\bV$.
\end{remark}

\begin{remark}
We may anticipate that $\beta_c$ will be a weaker marker of criticality than $\eta_c$ since it is not the nature of the temporal distribution of the degree of the $c$-th node that $\beta_c$ parametrises (unlike $\eta_c$), but it is the degree itself, of the $c$-th node of the RGG learnt with the full dataset ${\bf D}$. On the other hand, the sample estimate of the correlation matrix $\bSigma_b$ of data ${\bf D}_b$ - that is employed in the computation of $\eta_c$ - is likely to be a worse approximation of the inter-residue correlation matrix of this dataset $\forall b =1,\ldots,N_b$, than the sample estimate of $\bSigma$ is, of the correlation matrix of data ${\bf D}$. Such a statement stems from the smaller size ($N_b$ rows) of the sample used in estimating $\bSigma_b$, over the ($N_T$-sized) sample used to estimate $\bSigma$. Thus, reliability of $\eta_c$ may be less than that of $\beta_c$, $\forall c\in\bV$.
\end{remark}  
Details of our implementation of these three parameters towards the identification of the $n$-most critical residues of {\it{Skp1}}, using the simulated data ${\bf D}$, is presented in Section~\ref{sec:results}.

\subsection{Computing the inter-residue correlation matrix}
To learn the RGG variable, we need to compute the inter-column correlation matrix $\bSigma=[\rho_{i,j}]$ of the simulated dataset ${\bf D}$. Here $\rho_{i,j}$ is the value of the absolute correlation $\vert corr(X_i,X_j)\vert= \vert R_{i,j}\vert$ between the state of the $i$-th and that of the
$j$-th residues. Since a state variable (such as $X_i$) is a categorical variable - that takes values at eight levels in the data on {\it{Skp1}} - we need to compute $corr(X_i,X_j)$ using 
the Cram\'er's V measure of correlation between categorical variables \citep{cramer1946}.

\begin{definition}
  In dataset ${\bf D}$, to define the correlation between the categorical variables $X_i$ and $X_j$ - that take values at $N_{\ell}$-levels - we first construct an $N_{\ell}\times 2$-dimensional contingency table that we denote ${\cal T}_{i,j}$. The $\ell, q$-th cell of this table contains the frequency $\nu_{\ell, q}$ with which $X_q$ attains value at the $\ell$-th level, in data ${\bf D}$, $\forall q \in \{i,j\}$; $\ell\in\{1,\ldots,N_{\ell}\}$. 
Then $\chi^2_{i,j}$ is defined as: 
$\chi_{i, j}^2 := \displaystyle{\sum_{q=i,j}\sum_{\ell=1}^{N_{\ell}} \frac{(\nu_{\ell,q} - \mu_{\ell,q})^2}{\mu_{\ell.q}}},$
where $\mu_{\ell.q}$ is the mean of the frequency with which $X_q$ attains value in the $\ell$-th level in ${\bf D}$, (given by the product of the sum of the frequencies in the $\ell$-th row and $q$-th column, divided by the total frequency $N_{i,j}^{(T)}$ for $X_i$ and $X_j$ to attain values in all the $N_{\ell}$ levels, as displayed in the $N_{\ell}\times 2$-dimensional contingency table ${\cal T}_{i,j}$). Here $i < j; j\in\bV$. 
With this $\chi^{2}_{i,j}$, we construct
${\bSigma}=[\rho_{i,j}]$ s.t.:
\begin{equation}
  \label{eq:1}
    \rho_{i,j} = \sqrt{\frac{\chi^{2}_{i,j}/2 N_{i,j}^{(T)}}{min(N_{\ell}-1, 2-1)}}
    =  \sqrt{{\chi^{2}_{i,j}/2 N_{i,j}^{(T)}}},
\end{equation}
since there are $N_{i,j}^{(T)}$ number of realisations of $X_i$ and $X_j$ that we consider in the computation of this correlation.
Here $\rho_{i,j}$ is the value of the entry in the $i,j$-th cell of matrix $\bSigma$. Thus, the inter-column correlation matrix $\bSigma$ is computed for the dataset ${\bf D}$. We recall that in the simulated data on evolution of {\it{Skp1}}, $N_T= 10006$ and $N_{\ell}=8$. 
\end{definition}




\section{Results}
\label{sec:results}

\begin{figure}[H]
     \centering
     \begin{subfigure}
         \centering
         \includegraphics[width=0.35\textwidth]{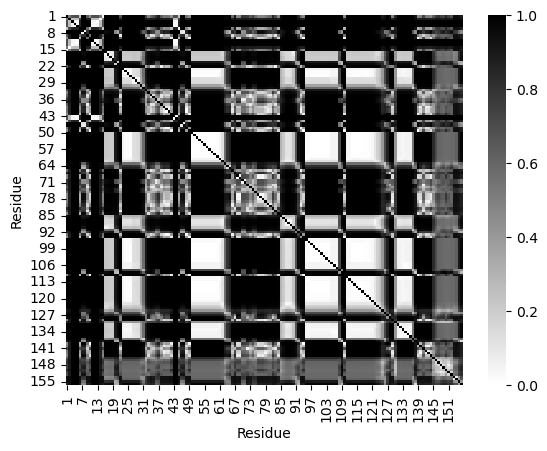}
     \end{subfigure}
     \begin{subfigure}
         \centering
         \includegraphics[width=0.45\textwidth]{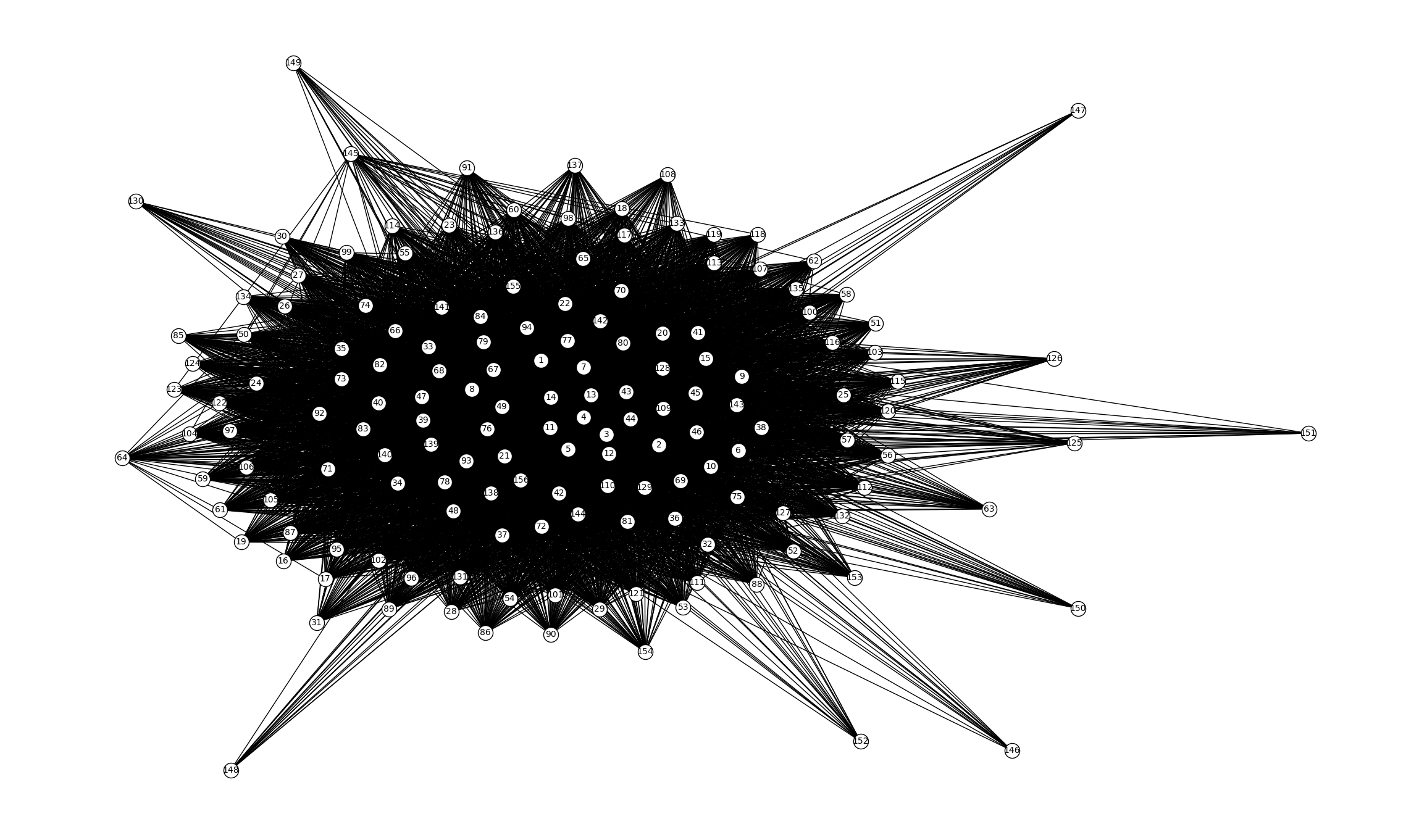}
\end{subfigure}
\caption{The heatmap generated by the correlation matrix $\bSigma$ is shown on the left. On the right, we present a learnt realisation of the RGG variable, given this correlation matrix $\bSigma$ of data ${\bf D}$, using a cut-off probability $\tau \approx 0.854$.}
\label{fig:heatmap_and_graph}
\end{figure}

To begin with, we calculate the inter-column correlation matrix $\bSigma$, of dataset ${\bf D}$, and learn the realisation of the RGG variable. These are presented in Figure~\ref{fig:heatmap_and_graph}.

\begin{figure}[H]
    \centering
    \includegraphics[width=0.4\textwidth]{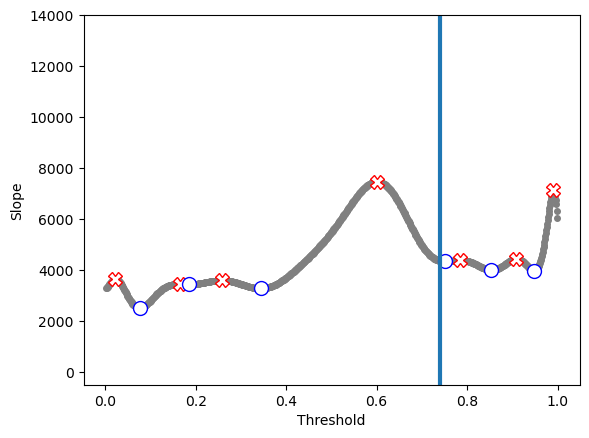}
    \caption{Figure displaying a fit to the computed (by differencing) values of (the slope with respect to $\tau$, or) the rate of change of the log posterior of the RGG variable, with changes in $\tau$, where the RGG is learnt using the simulated data ${\bf D}$ on states attained by residues of the protein {\it{Skp1}}. The vertical line indicates the ``best'' threshold value of $\omega\approx 0.740$ that is suggested by the COGENT method \citep{cogent_1}, implemented to learn the most consistent network in the data ${\bf D}$. The local maxima and minima of the slope values of the log posterior, are labelled in the unfilled crosses and circles respectively.}
\label{fig:diff_tau}
\end{figure}

\begin{figure}[H]
     \centering
     \begin{subfigure}
         \centering
         \includegraphics[width=0.3\textwidth]{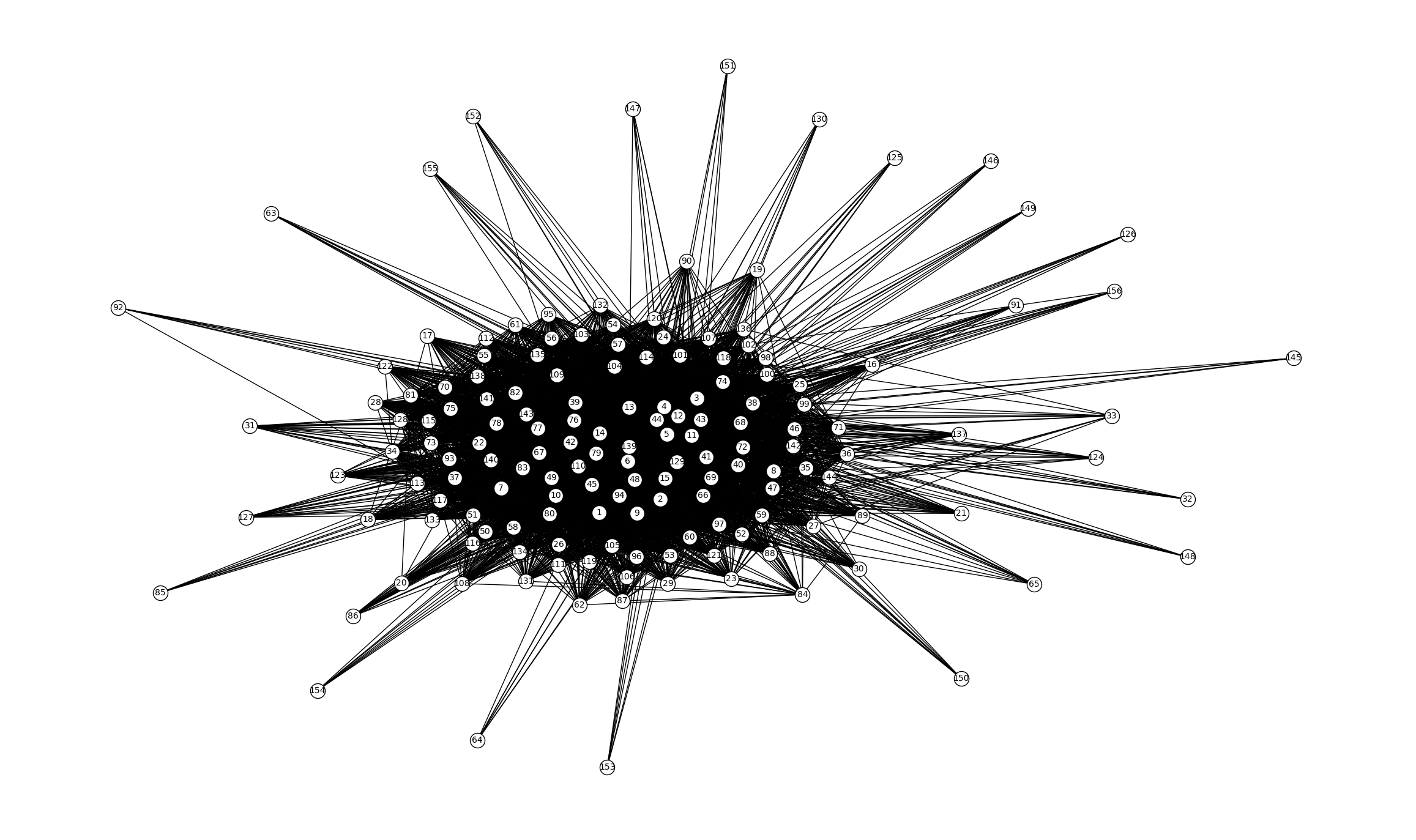}
     \end{subfigure}
     \begin{subfigure}
         \centering
         \includegraphics[width=0.3\textwidth]{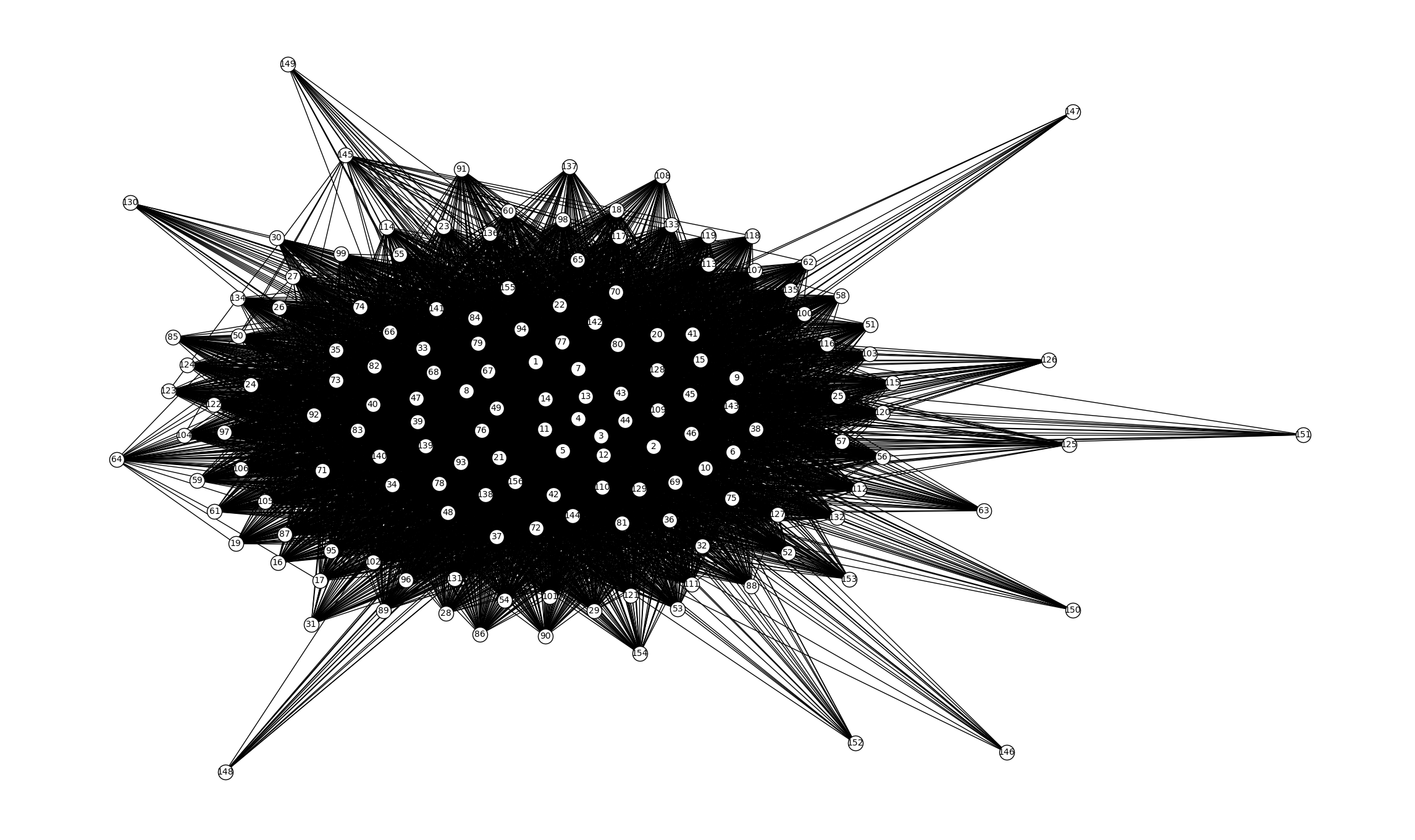}
\end{subfigure}
     \begin{subfigure}
         \centering
         \includegraphics[width=0.3\textwidth]{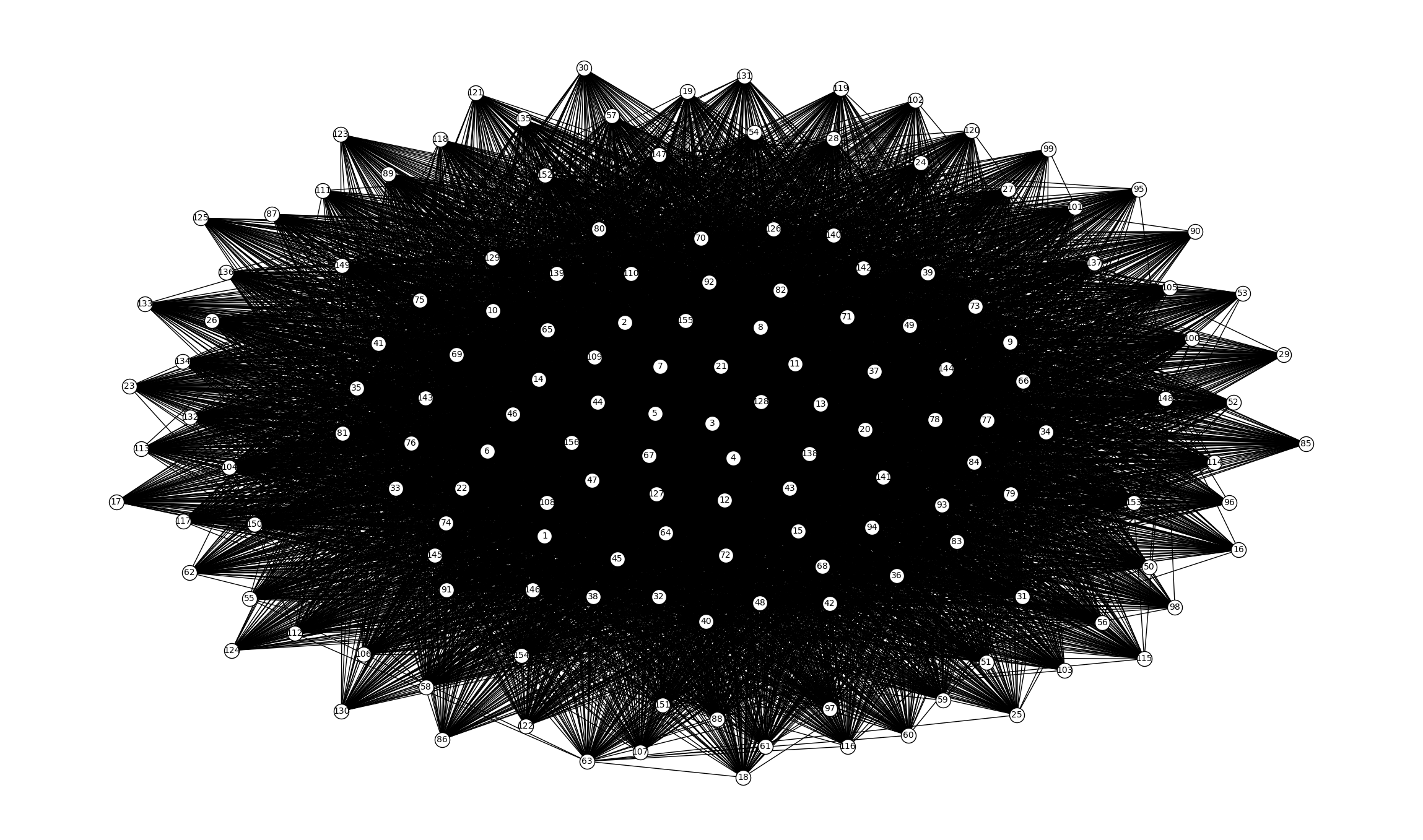}
\end{subfigure}
    \caption{Left: RGGs learnt for data ${\bf D}$, constructed at the COGENT-suggested threshold of $\omega \approx 0.740$.
    Middle: most robust RGG of data ${\bf D}$ realised at $\tau = \tau_{min}\approx$0.854.
    Right: most sensitive (to changes in $\tau$) RGG, at $\tau=\tau_{max}\approx$0.601.}
\label{fig:protein_graph}
\end{figure}

In Figure~\ref{fig:diff_tau}, we present results of the effect of the choice of the value of $\tau$ given the simulated data ${\bf D}$ of {\it{Skp1}}, while the effect of the COGENT-suggested
threshold $\omega$, \citep{cogent_1}, generated by the same data ${\bf D}$ is also presented. Additionally, in Figure~\ref{fig:protein_graph}, we display the most robust RGG that is learnt given data ${\bf D}$, at $\tau=\tau_{min}$, as well as the most-sensitive RGG learnt at $\tau=\tau_{max}$. Additionally, the figure includes the RGG that is learnt for this data, at the value of $\tau$ given by the thresholding method of \cite{cogent_1}. This threshold is the value $\tau\approx 0.740$ that produces the most consistent network, as per the COGENT thresholding technqiue.

\subsection{Criticality using $\delta$}
Our identification of the 20-most critical residues of the simulated protein {\it{Skp1}} is presented in 1st column of the table in Figure~\ref{all_crit_colour}. Here, the criterion used for this identification of criticality uses the computed $\delta_{\cdot}$ parameters (see Definition~\ref{defn:delta}). Thus, these 20-most critical residues are the 20-most functionally unstable residues in the simulated data ${\bf D}$.


\subsection{Criticality using $\eta$}
\begin{figure}[H]
     \centering
     \begin{subfigure}
         \centering
         \includegraphics[width=0.4\textwidth]{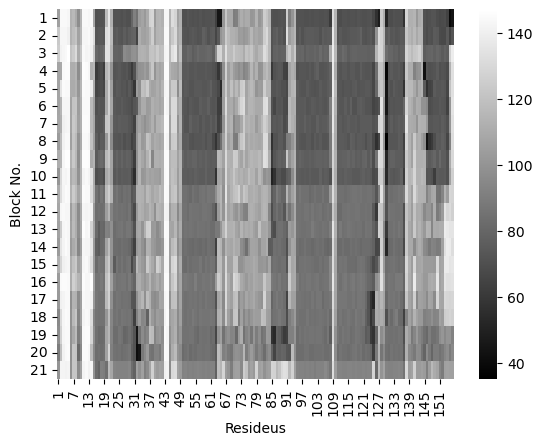}
     \end{subfigure}
     \begin{subfigure}
         \centering
         \includegraphics[width=0.4\textwidth]{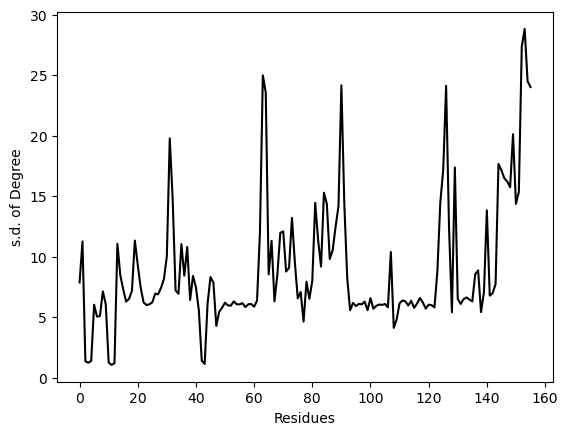}
\end{subfigure}
     \caption{Left: heat map representation of standard deviation of the sample of values of degree of $c$-th node, in each of the $N_b$ RGGs that are learnt at $\tau_{min}\approx 0.854$, given each $N_b$ (row-wise) partitions of data ${\bf D}$, plotted across all $c\in\{1,2,\ldots, 156\}$ and for all $N_b (=21)$ partitions. 
     Right: corresponding $\eta_c$ plotted against the residue index $c$, for $c=1,\ldots, 156$. } 
\label{fig:eta}
\end{figure}

We computed 
$\eta_c$ for $c=1,\ldots,p=156$, using the data on the residues, output from the simulation over times that pertain to ecach of the $N_b$ blocks. In Figure \ref{fig:eta} we plot the degree distribution against block index (along with a heat map representation). As clear from the heatmap, there are some isolated intervals of neighbouring residues that manifest high variation in the temporal distribution of degree. 

The column 2 of table in Figure~\ref{all_crit_colour}
shows the 20 most critical residues identified by computing $\eta_c$ parameter at $\tau \approx 0.854$, i.e. the 20 residues that exhibit maximal temporal infedilty in their interaction with other residues, during the evolution of {\it{Skp1}}.

\subsection{Criticality using $\beta$}
The coloumn 3 of the table in Figure~\ref{all_crit_colour}
shows the 20 most critical residues identified by the 20 smallest values of the degree distribution $\beta_c$ at $\tau \approx 0.854$, over the time during which the undertaken simulations offer output; (here $c=1,...,p=156$).

\subsection{Comparing identified criticality, with experimental results}
\label{sec:comparison}
\begin{figure}[H]
\centering
\includegraphics[width=0.8\textwidth]{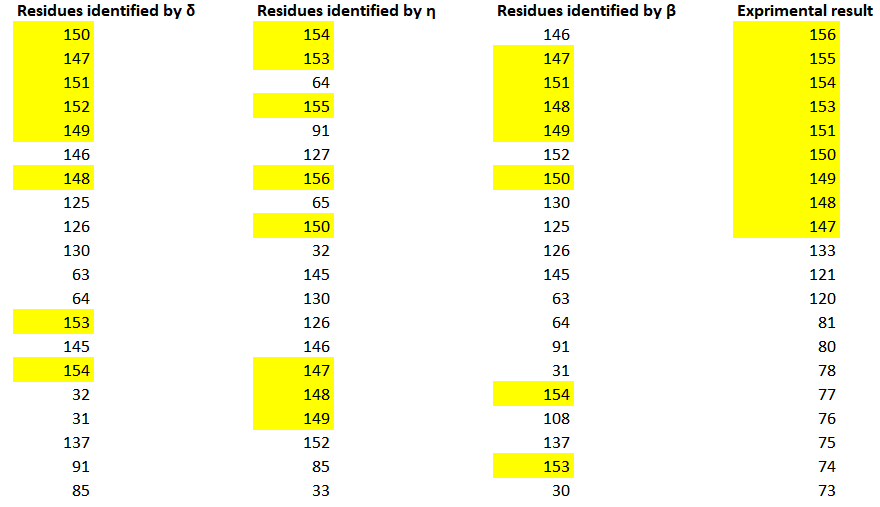}
\caption{Residues identified as critical, by $\delta_{\cdot}$, $\eta_{\cdot}$, $\beta_{\cdot}$ and experimental results. The shaded residues indicated themselves to be identified by all measures, given data ${\bf D}$.}
\label{all_crit_colour}
\end{figure}

\begin{figure}[H]
     \centering
     \includegraphics[width=\textwidth]{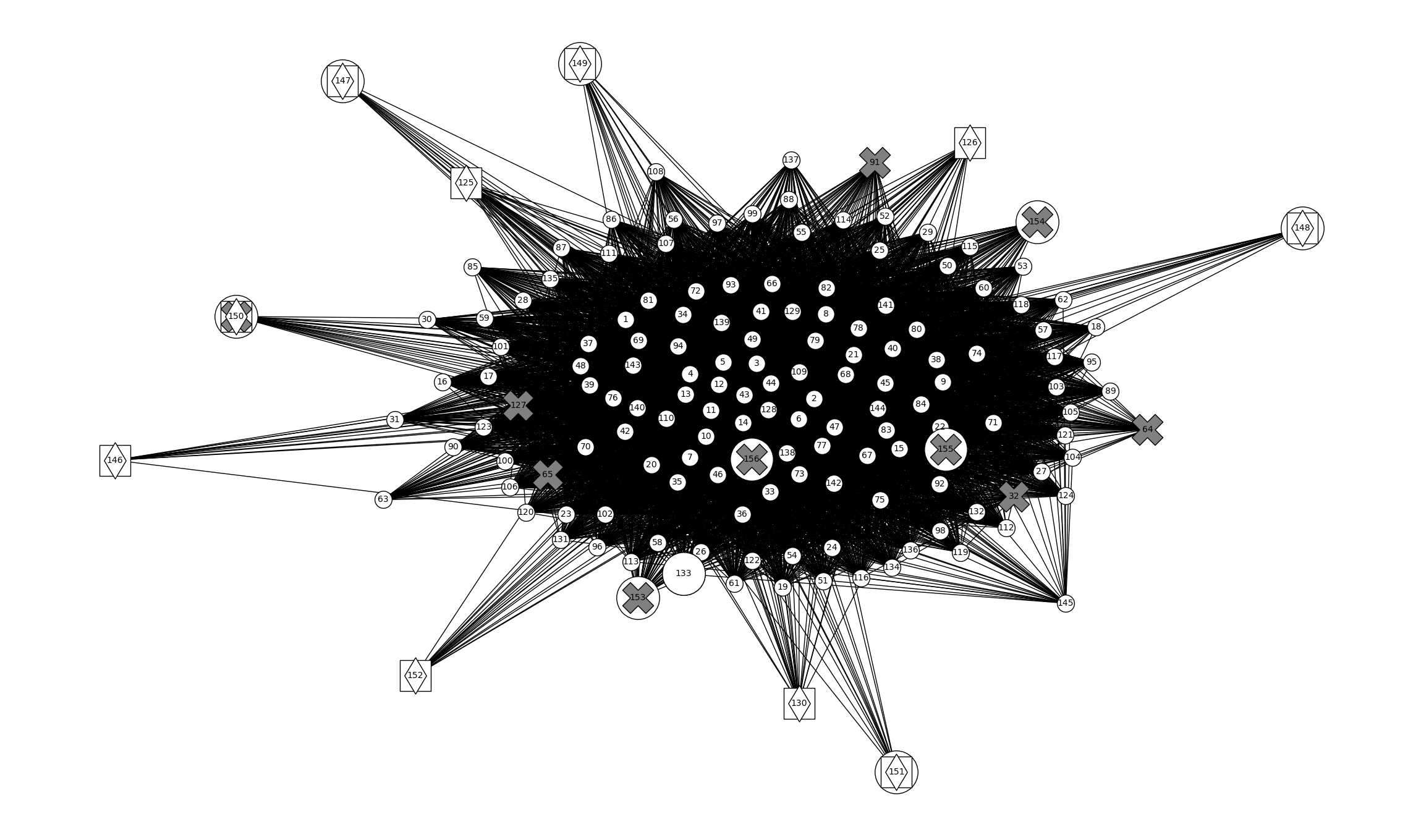}
\caption{10 most critical residues marked in yellow in Figure~\ref{all_crit_colour}, depicetd by $\delta_{\cdot}$ (as squares); $\eta_{\cdot}$ (grey crosses); $\beta_{\cdot}$ (diamonds); and experimental results (large white circles), marked on an RGG learnt of data ${\bf D}$.}
\label{all_crit}
\end{figure}

It can be seen that results of our identification of the most critical residues, with the three parameters $\delta_c, \eta_c, \beta_c$, enjoy some concurrence with each other, s.t. the residues that have the largest $\delta_{\cdot}$ values, are the residues that have the smallest $\beta_{\cdot}$ values. Again, residues that manifest maximal non-uniformity in their temporal degree distribution - i.e. their $\eta$ parameter - tally moderately well with the residues that have been identified as critical in our experimental results undertaken in the laboratory. These experiments identify criticality by tracking the frequency of change of state of the residues, where these experiments have the capacity to register changes that happen with frequency in [1/25, 1/2] milisecond $^{-1}$ approximately \cite{Reddy2018}. We tabulate these results below, in Figure~\ref{all_crit_colour}. In Figure~\ref{all_crit} we show the 10 most critical residues identified by computing the $\delta_{\cdot}$, $\eta_{\cdot}$, $\beta_{\cdot}$, and by the experiments, in an RGG of data ${\bf D}$ learnt at $\tau \approx 0.854$.



Out of the three parametrisations of criticality that we introduce above, it appears that identifying critical residues by computing the $\delta_{\cdot}$ parameter compares most favourably with the experimental results. The $\delta$ parameter maps to the physics inherent in the protein, without us having to invoke any model of the same, by informing on which residues contribute relatively more, towards the random graph variable that is learnt given the data on the evolution of the protein. We believe the $\delta$ parameter can override the challenges of a differently-shaped network/graph of distinct proteins, and provide a robust way of identifying functionally critical residues. 
We notice that our $\eta_{\cdot}$ parameter identifies a few critical residues which are also included as the 10-most critical residues in the experimental results, but fails to do this for the residues 151,149,148,147 that are experimentally identified as critical.
By design, the $\eta$ parameter can also handle clustered data - even if nodes group within multiple, isolated clusters in the graph of the full data, and critical residues are strewn across the different clusters. Abrupt initiation and termination of interactions of the $c$-th residue with other residues, will lead to comparatively higher unevenness in the temporal evolution of the degree of this node, than others. At the same time, we appreciate the sensitivity of the value of $\eta$ to the choice of the size of the block, given the time scale over which the inter-residual interactions happen, as recorded in the simulations. Then it is not surprising that an arbitrarily chosen partitioning did not yield the best results. Also, along with the $\beta_{\cdot}$ parameter, values of $\eta_{\cdot}$ are sensitive to the cut-off $\tau$ that we choose to draw the RGG of data ${\bf D}$ at. The three parameters that we devleop, have identified criticality beyond hitherto undertaken experiments, and mutagenesis experiments of these residues would be of future interest.

\end{document}